\documentclass[aps,prl,twocolumn,showpacs]{revtex4}
\usepackage{amssymb}
\usepackage{graphicx}
\usepackage{amsmath}
\usepackage{amsfonts}

\begin{document}

\title{Beyond Fermi-Liquid Theory: the $k$-Fermi liquids }
\date{\today}
\author{Tai-Kai Ng$^{1,2}$}
\affiliation{$^{1}$ Department of Physics, Hong Kong University of Science and Technology, Clear Water Bay, Hong Kong, China}
\affiliation{$^{2}$ Hong Kong Academy for Gifted Education, Shatin, Hong Kong, China}

\begin{abstract}
      We study in this paper the general properties of a many body system of fermions in arbitrary dimensions assuming that the {\em momentum} of individual fermions are good quantum numbers of the system. We call these systems $k$-Fermi liquids. We show how Landau Fermi liquid, Fermion liquid with exclusion statistics and spin-charge separation arises from this framework. Two exactly solvable $k$-Fermi liquid models, including the 1D Hubbard model and a toy model at arbitrary dimensions are discussed as examples of spin-charge separation.

\end{abstract}

\maketitle

\section{I: Introduction}
 A fundamental problem in quantum condensed matter physics is the identification of theoretical frameworks that describe plausible quantum phases of interacting many fermion systems. In this connection exact Bethe Ansatz\cite{bethe,bethe2} solution exists for a large class of Hamiltonian in one dimension (1D) because of the extra conservation laws 1D systems enjoy, and in higher dimensions a powerful framework that has proven to work for many interacting fermion systems is the Landau-Fermi liquid theory\cite{landau}. A common feature shared by the Bethe-Ansatz solutions and Landau Fermi liquid theory is that the eigenstates of the many-body Hamiltonian are characterized by a set of momenta $\{\mathbf{K}\}=\{ \vec{k}_1,\vec{k}_2,...,\vec{k}_N \}$ where $\vec{k}_i$'s represent the momentum of individual particles ($N=$ number of particles). For example, the low energy behavior for a system of spinless fermions is characterized by the effective energy\cite{landau}
  \begin{equation}
      \label{hlandau}
       E[\delta n]  = \sum_{\vec{k}}\varepsilon_{k}\delta n_{\vec{k}}+{1\over2V}\sum_{
      \vec{k};\vec{k}'}f_{\vec{k}\vec{k}'}\delta n_{\vec{k}}\delta n_{\vec{k}'}
      \end{equation}
    in Fermi liquid theory, where $\delta n_{\vec{k}}$ denotes the deviation of quasi-particle occupation number from ground state, $\varepsilon_{k}$ and $f_{\vec{k}\vec{k}'}$ are the quasi-particle energy and Landau interaction, respectively. The quasi-particles are labeled by their momentum $\vec{k}$'s which are good quantum numbers of the system.


   The Landau Fermi liquid theory was justified by a formal perturbation theory treating interaction between fermions as perturbation\cite{landau,aa}. Since 1980's, several new condensed matter physics systems were discovered that were believed to be outside the jurisdiction of Fermi liquid theory\cite{fqhe,gauge,ja}. To understand these systems theoretical frameworks that go beyond Fermi liquid states are needed\cite{ja,nm1}. In this paper we proceed along this direction by asking the question "what are the most general fermion liquid states that one can construct assuming that the set of momenta $\{ \vec{k}_1,\vec{k}_2,...,\vec{k}_N \}$ that describe the momentum of individual fermions (or quasi-particles) are good quantum numbers of the system"? We shall consider non-relativistic fermions both with and without spins in this paper. Our discussion is applicable to arbitrary dimensions.

 Before proceeding further, we first explain in more details the meaning of the statement that ``{\em the momentum of individual fermions $\{ \vec{k}_1,\vec{k}_2,...,\vec{k}_N \}$ are good quantum numbers of the system}". As illustration we consider spinless fermions with two-body, short range interaction $U(r)$ such that $U(r>R)=0$ ($r=$ distance between fermions and $R=$ range of interaction). We consider a low density fermion system such that the average distance between fermions $r_d$ is larger than $R$, In this case, there exists a finite region where the distance $r_{ij} (i\neq j=1,...,N)$ between fermions are all larger than $R$ and the behavior of the system is governed by the kinetic energy term only. The many-body wavefunction in this region is in general a superposition of Slater Determinants of plane wave states characterized by momentum of individual particles, $|\Psi>=\sum_la_l|\psi_{\{\mathbf{K}^l\}}>$, where $ \{\mathbf{K}^l\}=\{ \vec{k}^l_1,\vec{k}^l_2,...,\vec{k}^l_N \}$ denote sets of plausible momenta allowed by conservation laws and $|\psi_{\{\mathbf{K}^l\}}>$ is the Slater Determinant wavefunction constructed from the corresponding set of plane wave states $\{\mathbf{K}^l\}$.

  In one dimension, conservation of energy and momentum in two-body scattering implies that the set of momenta $\{\mathbf{K}\}$ remains unchanged during two-body scattering and suggests that the many-body eigenstates may consist of only one unique set of momenta $\{\mathbf{K}\}$. This observation forms the basis of Bethe-Ansatz solution\cite{bethe, bethe2}. In higher dimension the conservation of energy and momentum together with Pauli exclusion principle applies also to the {\em ground state + single quasi-particle excitation on the Fermi surface} and formed the starting point of Landau Fermi liquid theory\cite{landau}.
  The set of momenta $\{\mathbf{K}\}$ may stay unchanged also for the special case of interaction with forward-scattering only\cite{forward}. In this paper we shall not assume any specific form of interaction but shall assume phenomenologically that the many-body eigenstates consist of only one set of momenta $\{\mathbf{K}\}$ only and ask what are the plausible consequences of this assumption. A plausible wavefunction realization one may imagine for spinless fermions is a single slater determinant wavefunction characterized by a set of momenta $\{\mathbf{K}\}$ with a Jastrow factor that corrects for the short distance behavior of the wavefunction, i.e.
  \[ |\Psi(\{\mathbf{K}\})>=\prod_{i\neq j}g_{\{\mathbf{K}\}}(|\mathbf{r}_i-\mathbf{r}_j|)|\psi_{\{\mathbf{K}\}}(\mathbf{r}_1,\mathbf{r}_2,...,\mathbf{r}_N)>,  \]
  where $\mathbf{r}_i$ denotes the position of $i^{th}$ particle and $g_{\{\mathbf{K}\}}(r)$ is a $\{\mathbf{K}\}$-dependent Jastrow factor with $g_{\{\mathbf{K}\}}(r\rightarrow \infty)\rightarrow 1$.

   In our following discussions we shall not assume any particular microscopic Hamiltonian or wavefunctions besides global symmetry and conservation laws. We shall not assume that the quasi-particles are the quasi-particles in Landau Fermi liquid theory. For convenience we shall call the set of many-fermion states that are characterized by a fix set of momenta $\{\mathbf{K}\}$ the $k$-Fermi liquid states in the following. In section II we shall consider spinless fermions and see how exclusion statistics\cite{ex1, ex2} and Luttinger Theorem arises in this case. We note that in general for a fermion system with internal degrees of freedom the wave-function will not be completely specified by fixing only the set of momenta $\{\mathbf{K}\}$. A familiar case is the case of spin-$1/2$ fermions. We shall consider this situation in section III and see how Fermi liquid behavior\cite{landau} or spin-charge separation\cite{ja, hu1d,nz} may arise in this case. Our idea will be illustrated by two examples of exactly solvable spin-$1/2$ fermion models with spin-charge separation. A brief summary of the paper is given in section IV. A brief description of the properties of quasi-particles in general (spinless) $k$-Fermi liquids is given in the Appendix.

  \section{II: Spinless Fermions and exclusion statistics}
     Spinless fermions are the simplest example of $k$-Fermi liquids where the eigenstates can be characterized completely by the set of momenta $\{\mathbf{K}\}=\{ \vec{k}_1,\vec{k}_2,...,\vec{k}_N \}$ as there is no other degrees of freedom in the system. The only question that remains to be answered is: what determines the values of $\vec{k}_i$'s?

     To be specific, we consider a $d$-dimensional system of size $L^d$ with periodic boundary condition in all directions. We note that for interacting systems, only the {\em total} momentum $\vec{P}$ of the system is quantized by periodic boundary condition with
     \begin{equation}
     \label{ptot}
      \vec{P}=\sum_{i=1,N}\vec{k}_i={2\pi\over L}(n_1,.n_2,..n_d),
      \end{equation}
     where $n_i$ are integers. The individual values $\vec{k}_i$'s are determined by the details of the many-body wavefunction. Within the assumption of $k$-Fermi liquid, we may write
     \begin{equation}
     \label{kequation}
     \vec{k}_i=\vec{k}_{0i}+\vec{\kappa}_i(\{\mathbf{K}\}_i;U),
     \end{equation}
     where $\vec{k}_{0i}={2\pi\over L}(n_{1i},n_{2i},..,n_{di})$ is the single particle momentum for non-interacting fermions and $\vec{\kappa}_i$ denotes the correction to momentum coming from interaction\cite{bethe2,ex1,ex2}. $\vec{\kappa}_i$ is in general a function of all other occupied momenta $\vec{k}_j (j\neq i)$ (denoted by the set $\{\mathbf{K}\}_i$) and strength of interaction $U$. Eq.\ (\ref{kequation}) is in general an non-linear equation that determines self-consistently the allowed set of momenta $\{\mathbf{K}\}$. We shall assume that real value solutions $\{\mathbf{K}\}$ to the equation exists in our following discussions.

      The correction to momentum $\vec{k}_i$'s can be understood physically from the following consideration: we imagine trapping a fermion with momentum $\vec{k}_i$ to form a wave-packet centered at location $\vec{r}_0$ in our system, and then we move adiabatically the wave-packet along direction $\hat{x}$. Because of periodic boundary condition, the wave-packet returns to position $\vec{r}_0$ after traveling distance $L$. The many-body wavefunction returns to it's initial value except picking up a phase factor
      \begin{equation}
      \label{phase}
       k_{ix}L+\theta(\vec{k}_i;L\hat{x}+\vec{r}_0)-\theta(\vec{k}_i;\vec{r}_0)=2m\pi,
       \end{equation}
       the last equality is the requirement of periodic boundary condition. The phase factor $\Delta\theta(\vec{k}_i;L\hat{x})=\theta(\vec{k}_i;L\hat{x}+\vec{r}_0)-\theta(\vec{k}_i;\vec{r}_0)$ represents Berry phase corrections picked up by the wave-packet when it travels across the system. For a system with size $L>>r_d$ and with uniform density, we expect $\Delta\theta(\vec{k}_i;L\hat{x})\sim -L\kappa_{x}(\vec{k}_i)$ (thermodynamics limit) and
      \[ k_{ix}={2\pi m\over L}+\kappa_{ix},  \]
      which gives rise to Equation\ (\ref{kequation}) when we generalized the analysis to all directions.

       For 1D spinless fermions with short-range interaction, Bethe Ansatz solution gives rise to Eq.\ (\ref{kequation}) with a simple form of $\vec{\kappa}_i(\{\mathbf{K}\}_i;U)$ given by\cite{bethe2, sf1d}
        \begin{equation}
        \label{Bethe01}
        \vec{\kappa}_i(\{\mathbf{K}\}_i;U)=\sum_{\vec{k}_j\neq\vec{k}_i}\vec{\kappa}_{\vec{k}_i\vec{k}_j}(U).
        \end{equation}
         where $\vec{k}_j (\neq \vec{k}_i)$ denotes all other occupied momentum states. The system describes a fermion liquid with exclusion statistics as a result of the (non-uniform) shift in momenta $\vec{k}_{0}\rightarrow\vec{k}$\cite{bethe2,ex1,ex2}.

       \subsection{Adiabaticity, Luttinger Theorem and Fermi liquid}

       Using Eq.\ (\ref{kequation}), the total momentum of the $k$-Fermi liquid can be written as
       \begin{subequations}
       \label{ptot2}
     \begin{equation}
      \vec{P}=\sum_{i=1,N}\vec{k}_i=\vec{P}^{(0)}+\Delta\vec{P},
      \end{equation}
        where
        \begin{equation}
        \vec{P}^{(0)}=\sum_{i=1,N}\vec{k}_{0i}
        \end{equation}
        is the total momentum in the absence of interaction, and
        \begin{equation}
        \Delta\vec{P}=\sum_{i=1,N}\vec{\kappa}_i(\{\mathbf{K}\}_i;U)
        \end{equation}
        is the interaction-induced correction to total momentum.
        \end{subequations}

        We now consider turning on the interaction adiabatically from $U=0$. Since we assume translational invariance, the interaction term satisfies $[\vec{P},H_{int}]=0$ and the total momentum $\vec{P}$ should remains unchanged as interaction is turned on as long as the quantum many-body wavefunction is changing adiabatically. In this case, $\vec{P}=\vec{P}^{(0)}$ and
         \begin{equation}
         \label{pt2}
         \Delta\vec{P}=\sum_{i=1,N}\vec{\kappa}_i(\{\mathbf{K}\}_i;U)=0,
         \end{equation}
        which impose a constraint on the plausible forms of $\vec{\kappa}_i(\{\mathbf{K}\}_i;U)$. In the simple case of pairwise contribution\ (\ref{Bethe01}), we obtain
        \begin{subequations}
        \label{bequally}
        \begin{equation}
          \sum_{\vec{k}_i\vec{k}_j, i\neq j}\vec{\kappa}_{\vec{k}_i\vec{k}_j}(U)=0,
        \end{equation}
        or
        \begin{equation}
        \label{bconstraint}
        \vec{\kappa}_{\vec{k}_i\vec{k}_j}(U)=-\vec{\kappa}_{\vec{k}_j\vec{k}_i}(U)
        \end{equation}
        \end{subequations}
        which must be satisfied if the system consists of only two particles and remains valid for arbitrary number of particles as  $\vec{\kappa}_{\vec{k}_i\vec{k}_j}(U)$ is independent of other particles. The equality is satisfied by Bethe Ansatz solvable 1D spinless fermion models\cite{bethe2, sf1d} and expresses the physical requirement that for pairwise contributions, the change in momentum on particle $\vec{k}_i$ induced by particle $\vec{k}_j$ must be opposite to the change in momentum on particle $\vec{k}_j$ induced by particle $\vec{k}_i$ if the total momentum of the system remains conserved in the presence of interaction.

        We now apply this result to the ground state and low-energy excitations of spinless $k$-Fermi liquids assuming that the interacting $k$-Fermi liquid is connected adiabatically to the free fermion state, the free fermion ground state consists of a filled Fermi sea with Fermi momentum $\vec{k}_{0F}$ and total momentum $\vec{P}^{(0)}=0$. The corresponding interacting $k$-Fermi liquid has a filled Fermi sea with Fermi momentum
        \begin{equation}
        \label{kfermi}
        \vec{k}_F=\vec{k}_{0F}+\vec{\kappa}_{\vec{k}_F}(\{\mathbf{K}\}_{\vec{k}_F};U)
        \end{equation}
        and total momentum $\vec{P}=\vec{P}^{(0)}=0$.

       We now add a particle on the Fermi surface of the interacting $k$-Fermi liquid. It should be noted that the momentum of {\em all} other particles will be shifted according to Eq.\ (\ref{kequation}) when the particle with momentum $\vec{k}_F$ is added to the system\cite{ex2,exl} (see Appendix for a more detailed description). The total momentum carried by the excitation ($=\vec{k}_F +${\em change in momentum of all other particles in the system}) is
       \begin{equation}
       \label{lut1}
       \vec{P}=\sum_{i}\vec{k}_i=\vec{P}^{(0)}=\vec{k}_{0F}
       \end{equation}
        as a consequence of adiabaticity.

         We now apply this result to an excitation in a 1D $k$-Fermi liquid where a particle with Fermi momentum $-k_F$ is moved to momentum $k_F$ or vice versa. This is a zero-energy excitation with {\em total} momentum transfer $\pm 2k_{0F}$ because of adiabaticity as explained above. This result is in agreement with the proof of the Luttinger Theorem\cite{lut} by Yamanaka {et al.} in one dimension\cite{os1}, where they proved that a fermion system always carry zero energy excitation with momentum transfer $2k_{0F}$ if the system does not break translational symmetry. The argument can be extended rather straightforwardly to dimensions $D>1$. In this case we can construct zero energy excitations by moving fermions from one part of the Fermi surface with Fermi momentum $\vec{k}_F$ to another with Fermi momentum $\vec{p}_F$, the total momentum transfer being $\vec{p}_{0F}-\vec{k}_{0F}$, in agreement with Oshikawa's proof of Luttinger theorem at dimensions $D>1$\cite{os2}. It should be noted that Luttinger theorem does not measure the Fermi sea volume\cite{lut} enclosing the filled $\vec{k}$-points for general $k$-Fermi liquids.

         The low energy physics of a $k$-Fermi liquid is expected to be described by an effective Landau Fermi liquid type energy functional\ (\ref{hlandau}) as the eigenstates of these system are labeled completely by the particle momenta $\vec{k}$'s. However a quasi-particle in a $k$-Fermi liquid is not simply an independent particle with momentum $\vec{k}$ but the particle together with the shift in momentum of all other particles in the system. The total momentum carried by the quasi-particle is given by $\vec{p}=\vec{k}$+{\em change in momentum of all other particles in the system}$=\vec{k}_0$ (adiabaticity). The low energy properties of the system is thus described by a Landau Fermi liquid theory where the quasi-particles are labeled by their total momentum $\vec{p}$, with renormalized quasi-particle energy $\varepsilon_{k}\rightarrow\tilde{\varepsilon}_{p}$ and Landau interaction $f_{\vec{k}\vec{k}'}\rightarrow\tilde{f}_{\vec{P}\vec{P}'}$ (see Appendix for a brief description of the quasi-particle properties in spinless $k$-Fermi liquids). Notice that the Landau Fermi liquid theory satisfies Luttinger theorem when the quasi-particles are being labeled by the {\em total} momenta $\vec{p}$ they carry\cite{ex2,exl} because of Adiabaticity.

         A natural question that follows from the above discussions is: does the spectral function of the single-particle Green's function exhibits a peak at momentum $\vec{k}$ or at total momentum $\vec{p}=\vec{k}_{0}$? 

         We argue that the spectral function exhibits a peak at total momentum $\vec{p}$. The argument is based on Eq.\ (\ref{phase}) assuming that the equation is applicable for wave-packet traveling distance $L>X>>r_d$, i.e., thermodynamics limit. In this case the phase picked up by the many-body wavefunction after the wave-packet travels distance $X\hat{x}$ is given by
    \[
       k_{ix}X+\theta(\vec{k}_i;X\hat{x}+\vec{r}_0)-\theta(\vec{k}_i;\vec{r}_0)\sim 2m\pi\left({X\over L}\right)=k_{0x}X.  \]

       Generalizing to all directions, we find that the total phase pick up by the wave-packet after traveling distance $\vec{X}$ (which is what is measured by the single-particle Green's function) is $\sim \vec{k}_{0}.\vec{X}$, suggesting that the single-particle Green's function has a peak at total momentum $\vec{p}=\vec{k}_0$ but not $\vec{k}$, in agreement with Fermi liquid theory.

        \subsection{Beyond simple Fermi liquids}
        Let's imagine a spinless $k$-Fermi liquid that is adiabatically connected to non-interacting fermions for interaction strength $U$ satisfying $U_1>U>0$. A phase transition occurs at $U_1$ and the state enters a new quantum state $|S\rangle$. The system stays at state $|S\rangle$ and evolves adiabatically for $U_2\geq U\geq U_1$. To be more specific we assume that continuous translational symmetry is broken and the system only satisfies a discrete (lattice) translation symmetry in the state $|S\rangle$. In this case, the system may still be described by a $k$-Fermi liquid state except that the $\vec{k}$-vectors now describe the crystal momentum and are restricted to the first Brillouin zone. What else can we say about the system in this case?

         Let the state be specified by the set of momenta $\{\mathbf{K}^{(1)}\}$ at interaction strength $U_1$.  We may write
         \begin{equation}
        \label{kequationu}
         \vec{k}_i=\vec{k}_{1i}+\vec{\kappa}_i^{(1)}(\{\mathbf{K}\}_i;U-U_1),
         \end{equation}
         for $U_2\geq U\geq U_1$, where $\vec{k}_{1i}\in\{\mathbf{K}^{(1)}\}$ is the single particle momentum for $i^{th}$ particle at $U=U_1$ and $\vec{\kappa}_i^{(1)}(\{\mathbf{K}\}_i;0)=0$. One may repeat the analysis at previous section to show that adiabaticity implies that the total momentum satisfies $\vec{P}(U)=\vec{P}(U_1)+\vec{P}_Q$ for $U_2\geq U\geq U_1$ where $\vec{P}_Q$ is a reciprocal lattice vector (including $\vec{P}_Q=0$). Correspondingly, Eq.\ (\ref{pt2}) becomes
         \begin{subequations}
          \begin{equation}
          \label{pt2g}
         \vec{P}=\vec{P}^{(1)}+\Delta\vec{P}=\sum_{i}\vec{k}_{1i}+\sum_{i}\vec{\kappa}^{(1)}_i(\{\mathbf{K}\}_i;U-U_1)
         \end{equation}
         with
         \begin{equation}
        \sum_{i}\vec{\kappa}^{(1)}_i(\{\mathbf{K}\}_i;U-U_1)=\vec{P}_Q.
        \end{equation}
        \end{subequations}

          Next we consider adding a quasi-particle with momentum $\vec{k}_i$ on the system. The total momentum carried by the quasi-particle is
          \[
             \vec{p}_i=\vec{k}_{i}+\sum_{j\neq i}\vec{\kappa}^{(1)}_{\vec{k}_j}(\{\mathbf{K}\}_j;U-U_1)=\vec{k}_{1i}+\vec{P}_Q,  \]
             following previous arguments.

          Physically, it is expected that $\Delta\vec{P}=\vec{P}_Q\neq0$ only when $\vec{k}_i$ is at the Brillouin zone boundary so that $\varepsilon_{\vec{k}_{i}}=\varepsilon_{\vec{k}_{i}+\vec{P}_Q}$, i.e. Umklapp processes. In this case the Fermi surface remains independent of $U$ at regions where the Umklapp processes are unimportant if the Fermi surface is defined by the {\em total} (crystal) momentum carried by the quasi-particles. Exceptions are found at regions close to the Brillouin zone boundary, the details of these changes depend on the microscopic Hamiltonian and are outside the scope of this paper.

          The low energy physics of the $k$-Fermi liquid states are described by Landau Fermi liquid theory in this regime as in the case with $U_1>U>0$ except that the total momentum $\vec{p}$ carried by a quasi-particle in phase $|S\rangle$ is in general not equal to $\vec{k}_0$ because $|S\rangle$ is not adiabatically connected to the non-interacting fermion state and Luttinger Theorem is in general not satisfied.

        \section{spin-$1/2$ fermions: from Landau Fermi liquids to spin-charge separation}

         The $k$-Fermi liquids with internal degrees of freedom are much more interesting than the simple case of spinless fermions because the momentum set $\{\mathbf{K} \}$ does not characterize all degrees of freedom in the system. Here we shall consider spin-$1/2$ fermions as example. We shall assume that our systems have both translational and spin-rotational invariance (and no spin-orbit coupling) such that the total momentum $\vec{P}$ and total spin $\vec{S}$ are conserved. To start with, we consider the simple case when the $k$-Fermi liquid is adiabatically connected to non-interacting, spin-$1/2$ fermions.

         \subsection{Fermi liquid state}
            A straightforward way to extend the spinless $k$-Fermi liquid phenomenology to spin-$1/2$ fermions is to replace Eq.\ (\ref{kequation}) by
         \begin{equation}
       \label{kequation2}
       \vec{k}_{i\sigma}=\vec{k}_{0i\sigma}+\vec{\kappa}_{i\sigma}(\{\mathbf{K}\mathbf{\sigma}\}_{i\sigma};U),
       \end{equation}
           where we label the individual fermion states by both momentum $\vec{k}$ and spin $\sigma=\uparrow,\downarrow$. Notice we have assumed that both the momentum $\vec{k}_i$ and spin $\sigma_i$ are {\em good} quantum numbers in writing down Eq.\ (\ref{kequation2}), which is the case for non-interacting fermi gas.  The eigenstates of the $k$-Fermi liquid are characterized by the spin and momentum of individual single particle states $\{\mathbf{K}\mathbf{\sigma}\}= \{\vec{k}_1\sigma_1,\vec{k}_2\sigma_2,...,\vec{k}_N\sigma_N\}$ and $\vec{\kappa}_{i\sigma}(\{\mathbf{K}\mathbf{\sigma}\}_{i\sigma};U)$ is a function of all occupied fermion states ($\vec{k}_j\sigma_j$) with $j\neq i$ (denoted by $\{\mathbf{K}\mathbf{\sigma}\}_{i\sigma}$).

        For $k$-Fermi liquids that are adiabatically connected to non-interacting fermions, we obtain
        \[
        \sum_{i\sigma}\vec{\kappa}_{i\sigma}(\{\mathbf{K}\mathbf{\sigma}\}_{i\sigma};U)=0,  \]
        as a result of momentum conservation which implies for pairwise contribution
        \begin{equation}
        \label{Bethe1}
        \vec{\kappa}_{i\sigma}(\{\mathbf{K}\mathbf{\sigma}\}_{i\sigma};U)=\sum_{j\sigma'(\neq i\sigma)}\vec{\kappa}_{\vec{k}_i\sigma\vec{k}_j\sigma'}(U),
        \end{equation}
        with
        \[ \vec{\kappa}_{\vec{k}\sigma\vec{k}'\sigma'}(U)=-\vec{\kappa}_{\vec{k}'\sigma'\vec{k}\sigma}(U),  \]
        similar to the case of spinless fermions.

        The Luttinger Theorem can be proved in a similar way as for spinless fermions and the effective low energy theory describing the $k$-Fermi liquid states is a Landau Fermi liquid theory for spin-$1/2$ fermions when the fermion excitations are labeled by {\em total spin and momentum} they carry. Elementary excitations are constructed by adding/removing $k$-fermions which carry both spin and charge. 

          We note that the momentum-defining equation\ (\ref{kequation2}) gives rise to the same set of $\vec{k}_{i\uparrow}$ and $\vec{k}_{i\downarrow}$ if $\vec{k}_{i\uparrow}^{(0)}=\vec{k}_{i\downarrow}^{(0)}$ $\forall i$ and the function $\vec{\kappa}_{i\sigma}(\{\mathbf{K}\sigma\};U)$ respects spin-rotation symmetry. This happens when there are equal number of $\uparrow$- and $\downarrow$-spin fermions occupying the same set of single momentum states. The degeneracy between $\vec{k}_{i\uparrow}$ and $\vec{k}_{i\downarrow}$ is the hallmark of a non-magnetic Fermi liquid ground state where both $\uparrow$- and $\downarrow$-spin fermions occupy the same set of momenta $\{\mathbf{K}\}$. We shall revisit this issue when we consider spin-charge separation.

        Lastly we note that a Fermi-liquid like $k$-Fermi liquid state with fixed $M$ spin-up particles and $N-M$ spin-down particles is characterized by the $M$ momenta of spin-up particles and $N-M$ momenta of spin-down particles. Thus the state is specified by $M+(N-M)=N$ quantum numbers. There is no extra quantum number associated with the spin-degrees of freedom.

        \subsection{General spin-$1/2$ $k$-Fermi liquids}

      The situation becomes more complicated if we do not assume Fermi-liquid adiabaticity and allow more general spin configurations associated with a (fixed) momenta configuration $\{\mathbf{K}\}$. For example, for a system with two fermions with opposite spins occupying the momentum states $\vec{k}_1, \vec{k}_2$, the four configurations
     \begin{eqnarray}
     \label{wfs}
     \psi_{(\vec{k}_1\uparrow)(\vec{k}_2\downarrow)}(\vec{r}_1,\vec{r}_2) & = & e^{i[\vec{k}_1.\vec{r}_1+\vec{k}_2.\vec{r}_2]}|\uparrow_1\downarrow_2>
      - (\vec{r}_1\rightleftarrows\vec{r}_2), \\  \nonumber
     \psi_{(\vec{k}_1\downarrow)(\vec{k}_2\uparrow)}(\vec{r}_1,\vec{r}_2) & = & e^{i[\vec{k}_1.\vec{r}_1+\vec{k}_2.\vec{r}_2]}|\downarrow_1\uparrow_2>
      - (\vec{r}_1\rightleftarrows\vec{r}_2), \\  \nonumber
     \psi^{\pm}_{(\vec{k}_1\vec{k}_2)(\uparrow\downarrow))}(\vec{r}_1,\vec{r}_2) & =  & \left(e^{i[\vec{k}_1.\vec{r}_1+\vec{k}_2.\vec{r}_2]}\pm e^{i[\vec{k}_1.\vec{r}_2+\vec{k}_2.\vec{r}_1]}\right)  \\ \nonumber
      & & \times \left(|\uparrow_{1}\downarrow_{2}>\mp|\uparrow_{2}\downarrow_{1}>\right)
      \end{eqnarray}
     represent four different quantum states of the system, and only the first two are being used in constructing Fermi-liquid states.

     When more general spin-configurations are allowed, the corresponding single particle momentum defining equation should take the form
       \begin{equation}
      \label{kequation3}
      \vec{k}_{i}=\vec{k}_{0i}+\vec{\kappa}_{i}(\{\mathbf{K};\mathbf{Q}\}_i;U),
      \end{equation}
       where $\{\mathbf{K}\}$ specifies the set of $k$-values occupied by the fermions and $\{\mathbf{Q}\}$ is a set of quantum numbers specifying the quantum spin state. We don't assign both spin and  momentum to a single particle state and the momenta set $\{\mathbf{K}\}$ contains $N$ different momentum values $=$ number of fermions in the system in general.

        For a given set of momenta $\{\mathbf{K}\}$, the spin state $\{\mathbf{Q}\}$ is in general determined by an effective quantum spin model with $N$ spins defined on $N$ momentum points $\vec{k}_i (i=1,..,N)\in\{\mathbf{K}\}$, i.e. a {\em $N$-site quantum spin model defined in momentum-space  $\{\mathbf{K}\}$}.  An example of effective spin Hamiltonian in $k$-space is
       \begin{subequations}
       \label{heffspin}
       \begin{equation}
       H_{eff}^S={1\over 2V}\sum_{\vec{k}\neq\vec{k}'\in\{\mathbf{K}\}}F_{\vec{k}\vec{k}'}\vec{S}_{\vec{k}}.\vec{S}_{\vec{k}'}
       \end{equation}
       where
       \begin{equation}
       \vec{S}_{\vec{k}} = \Psi_{\vec{k}}^+\mathbf{\vec{\sigma}}\Psi_{\vec{k}}.
       \end{equation}
       $\Psi^+_{\vec{k}} = (c^+_{\vec{k}\uparrow}\ c^+_{\vec{k}\downarrow})$ are spinors in $k$-space and $\mathbf{\vec{\sigma}}$ are Pauli matrices.
       \end{subequations}
       The ground state of the $k$-space spin Hamiltonian may describe a spin-ordered state or a spin-liquid state in $k$-space.

       Let $M$ be the number of down spins in a $k$-Fermi liquid system with $N$ fermions. Let $\bar{M}=\min(M,N-M)$, the spin wave-function is then specified by $\bar{M}$ variables which mark the positions of the minority spin species in momentum space $\{\mathbf{K}\}$ and the spin-eigenstates are specified by $\bar{M}$ quantum numbers. The total number of variables specifying a general spin-$1/2$ $k$-Fermi liquid state is therefore $N$ (which specifies the $\vec{k}$'s) $+ \bar{M}$ (which specifies the spin state), which is larger than the corresponding quantum number specifying a Fermi liquid state. This is exactly the situation of one-dimensional Hubbard model, where the eigenstates are specify by $N$ charge momenta + $\bar{M}$ spin rapidities\cite{hu1d}.

       \subsubsection{Fermi liquid - non-Fermi liquid transition }

         Let's consider an isotropic $k$-Fermi liquid with $2N$ spin-$1/2$ fermions in a non-magnetic Fermi liquid ground state. We define here a Fermi liquid ground state as a state where the 2N fermions occupying N lowest momentum states $\vec{k}_{i}$. In this case there is only one possible spin configuration where each momentum $k_i$ carries a pair of spins $(\uparrow,\downarrow)$. This can be seen easily from Eq.\ (\ref{wfs}) where the four wavefunctions collapse to only one spin-singlet configuration when $\vec{k}_1=\vec{k}_2$ because of the Pauli exclusion principle.

        A state with extra spin-degree of freedom is possible when the $2N$ spin-$1/2$ fermions are allowed to occupy more than $N$ different momentum states $\vec{k}_i$. As an example we consider a four fermion system occupying two momentum states $\vec{k}_1,\vec{k}_2$. The only possible fermion quantum state that can be formed in this case is a Fermi liquid state where each momentum state is occupied by 2 fermions forming a spin-singlet. On the other hand, if the four fermion system occupies three (four) different momentum states $\vec{k}_1,\vec{k}_2,\vec{k}_3$, ($\vec{k}_4$), it is easy to show that there exists a total of six (twelve) independent spin states where one may build a quantum ground state on. The existence of more than one possible spin configurations for fixed set of ground state momentum vectors $\{\mathbf{K}\}$ provides a necessary condition for spin-charge separation in $k$-Fermi liquids where one may create spin-excitations without changing the momentum configuration $\{\mathbf{K}\}$ of the fermions. (Notice however that the {\em values} of $\vec{k}_i$'s may depend on the quantum spin configuration in general $k$-Fermi liquids as a result of exclusion statistics).

         This analysis suggests that a $k$-Fermi liquid with $2N$ fermions may undergo a {\em Fermi-liquid - non-Fermi liquid (with spin-charge separation)} transition driven by a change in the number of independent particle momenta in the ground state $N_k$ from $N_k=N$ to $N_k>N$ because of spin-spin interaction\ (\ref{heffspin}). We note that the spin-charge separated state is not adiabatically continued to the Fermi liquid state because of the change in $N_k$.

       \subsection{Examples of $k$-Fermi liquids with spin-charge separation}

        We now present two exactly solvable models representing $k$-Fermi liquids with spin-charge separation as defined in last sub-section. To illustrate the idea of Fermi liquid - non-Fermi liquid (with spin-charge separation) transition we first consider a toy model that exhibits the transition with no exclusion statistics effect, i.e. $\vec{\kappa}_{i}(\{\mathbf{K};\mathbf{Q}\}_i;U)=0$. We then consider the one-dimensional Hubbard model which is a model exhibiting both exclusion statistics and spin-charge separation\cite{bethe2,hu1d}.

        \subsubsection{$k$-Space Valence Bond model}
        We consider a lattice model Hamiltonian
        \begin{equation}
        \label{kvbm}
        H=\sum_{\vec{k}\sigma}\xi_{\vec{k}}c^+_{\vec{k}\sigma}c_{\vec{k}\sigma}+{1\over2}\sum_{k}J_{\vec{k}}\vec{S}_{\vec{k}}.\vec{S}_{-\vec{k}}
        \end{equation}
        where $\vec{S}_{\vec{k}}$ is the spin operator defined in Eq.\ (\ref{heffspin}). The first part of the Hamiltonian is the usual kinetic energy term for spin-$1/2$ fermions, whereas the second term describes a Heisenberg interaction between electrons occupying opposite momentum states. $\vec{k}$'s are the allowed momenta defined on the corresponding non-interacting lattice model. We shall assume $J_{\vec{k}}=J_{-\vec{k}}\geq0$ and $\xi_{\vec{k}}=\xi_{-\vec{k}}$ in the following. The Hamiltonian can be rewritten as a sum of independent momentum components
        \begin{subequations}
        \label{kvbs2}
        \begin{equation}
        H ={1\over2}\sum_{\vec{k}}H_{\vec{k}},
        \end{equation}
        where
        \begin{equation}
        H_{\vec{k}} = \sum_{\sigma}\xi_{\vec{k}}(c^+_{\vec{k}\sigma}c_{\vec{k}\sigma}+c^+_{-\vec{k}\sigma}c_{-\vec{k}\sigma})
        +J_{\vec{k}}\vec{S}_{\vec{k}}.\vec{S}_{-\vec{k}}
        \end{equation}
        \end{subequations}
         can be diagonalized easily. The eigenstates of $H_{\vec{k}}$ can be divided into sectors with different number of fermions $n$ with eigen-energies $E_n$ and degeneracy $d_n$,
        \begin{eqnarray}
        \label{hkd}
        n=0, & &  E_0=0,  d_0 = 1  \\ \nonumber
        n=1, & &  E_1=\xi_{\vec{k}},  d_1 =4  \\ \nonumber
        n=2, & &  E_2^{(1)}=2\xi_{\vec{k}}, d_2^{(1)}=2   \\ \nonumber
             & &  E_2^{(s)}=2\xi_{\vec{k}}-{3J_{\vec{k}}\over4}, d_2^{(s)}=1   \\ \nonumber
             & &  E_2^{(t)}= 2\xi_{\vec{k}}+{J_{\vec{k}}\over4}, d_2^{(t)}=3   \\ \nonumber
        n=3  & &  E_3=3\xi_{\vec{k}}, d_3=4  \\ \nonumber
        n=4  & &  E_4=4\xi_{\vec{k}}, d_4=1.
        \end{eqnarray}

        We have set $\hbar=1$ in presenting our results. There are altogether $2^4=16$ possible states in each $\vec{k}$-sector, and the Heisenberg interaction is effective only in the $n=2$ sector with one fermion occupying state $\vec{k}$ and another occupying $-\vec{k}$. The two fermions may form a spin singlet with energy $E_2^{(s)}$ or a spin triplet with energy $E_2^{(t)}$ in this case.

       It is easy to see that the ground state of $H_{\vec{k}}$ has $n=4$ when $2\xi_{\vec{k}}<-{3J_{\vec{k}}\over4}$, and has $n=0$ when $2\xi_{\vec{k}}>{3J_{\vec{k}}\over4}$. It is in the $n=2$ spin-singlet sector when ${3J_{\vec{k}}\over4}>2|\xi_{\vec{k}}|$.

        The fermion occupation number $n_{\vec{k}\sigma}$ in the ground state thus has $n_{\vec{k}\uparrow}=n_{\vec{k}\downarrow}=n_{\vec{k}}$ with the following features:
        \begin{eqnarray}
        \label{nkvb}
           n_{\vec{k}}=1, & & 2\xi_{\vec{k}}<-{3J_{\vec{k}}\over4}  \\ \nonumber
           n_{\vec{k}}={1\over2}, & & {3J_{\vec{k}}\over4}>2|\xi_{\vec{k}}|  \\ \nonumber
           n_{\vec{k}}=0, & & 2\xi_{\vec{k}}>{3J_{\vec{k}}\over4}
           \end{eqnarray}
         We shall call this the $k$-Space Valence Bond (kVB) state in the following, since the state consists of fixed valence bond (or spin-singlet) pairs between opposite momentum states in regions satisfying ${3J_{\vec{k}}\over4}>2|\xi_{\vec{k}}|$. The state has spin-charge separation according to our definition, since we may excite a valence bond pair to become a spin triplet without affecting the occupied momenta $\vec{k}$'s, the spin excitation energy is $J_{\vec{k}}$. In particular, a quantum phase transition from a Fermi liquid state ($N_k=N$) to the kVB state ($N_k>N$) occurs if the interaction strength $J_{\vec{k}}$ changes from ${3J_{\vec{k}}\over4}<2|\xi_{\vec{k}}|$ $\forall \vec{k}$ to ${3J_{\vec{k}}\over4}>2|\xi_{\vec{k}}|$ at some regions of momentum $\vec{k}$.

           We next consider excitations in the kVB state. It is straightforward to show that single particle excitations where a fermion is added or removed from the system are gapped with minimum excitation energy ${3J_{\vec{k}}\over8}$. Particle-hole excitations constructed by moving a fermion from an occupied momentum state $\vec{k}$ to an unoccupied state $\vec{k}'$ costs minimum energy ${3J_{\vec{k}}\over8}+{3J_{\vec{k}'}\over8}$ because our toy Hamiltonian\ (\ref{kvbm}) is divided into independent $\vec{k}$-sectors. Thus the energy of moving a fermion from one momentum state to another is simply the sum of the two single-particle contributions. We note that these excitations become gapless if $J_{\vec{k}}\rightarrow0$ at some parts of Fermi surface.

           The one-particle Green's function can be computed straightforwardly using the spectral representation. We obtain at temperature $T\rightarrow0$,
           \begin{subequations}
       \label{green0}
        \begin{equation}
      G(\vec{k},i\omega)\rightarrow {1 \over i\omega-\xi_{\vec{k}}},
       \end{equation}
       for ${3J_{\vec{k}}\over4}<2|\xi_{\vec{k}}|$ and
       \begin{equation}
      G(\vec{k},i\omega)\rightarrow {1\over 2}\left({1\over i\omega-{3J_{\vec{k}}\over 4}-\xi_{\vec{k}}}+{1\over i\omega+{3J_{\vec{k}}\over 4}-\xi_{\vec{k}}}\right)
      \end{equation}
      for ${3J_{\vec{k}}\over4}>2|\xi_{\vec{k}}|$. The first term comes from $E_2\rightarrow E_3$ (particle) excitation and the second term comes from $E_2\rightarrow E_1$ (hole) excitation.
      \end{subequations}

         The finite temperature ($T\neq0$) properties of the system can also be computed easily. it is straightforward to show that the Partition function $Z(\beta)$ ($\beta=(k_BT)^{-1})$ is given by $Z(\beta)=\sqrt{\prod_{\vec{k}}z(\beta,\vec{k})}$, where
     \begin{eqnarray}
     \label{partk}
     z(\beta,\vec{k}) & = & z(\beta,-\vec{k})  \\ \nonumber
     & = & \left(1+4x_k+x_k^2(2+y_k^3+3y_k^{-1})+4x_k^3+x_k^4\right),
     \end{eqnarray}
     where $x_k=e^{-\beta\xi_{\vec{k}}}$ and $y_k=e^{\beta J_{\vec{k}}/4}$.


      It is interesting to note that although single-particle and particle-hole excitations are gapped in our model, it supports gapless charge $=2$, spinless excitations even if $J_{\vec{k}}>0$ $\forall\vec{k}$. The excitation can be formed by removing a spin-singlet pair from the top of the $n_{\vec{k}}=1/2$ region or by adding a spin singlet pair at the bottom of the $n_{\vec{k}}=0$ region. The excitation energies in these two cases are $\Delta E=-(+)(2\xi_{\vec{k}}-{3J_{\vec{k}}\over4})\rightarrow0$ at the edge region $2\xi_{\vec{k}}\rightarrow{3J_{\vec{k}}\over4}$. Similarly it is easy to show that gapless excitations can be formed by adding two particles at the bottom of the $n_{\vec{k}}=1/2$ region or by removing two particles leaving a spin-singlet pair at the top of the $n_{\vec{k}}=1$ region. It can be shown using Eq.\ (\ref{partk}) that the gapless charge-2 excitations leads to a low temperature specific heat $\sim\gamma T$ in our toy model.

      The system is thus a charge-$2$ metal with spin-gap. The charge-$2$ singlets may be identified as Cooper pairs and the low energy sector of our toy model describes a strongly quantum disordered BCS superconductor (or a Cooper pair metal) where quantum phase coherence between the Cooper pairs is completely lost\cite{fisher, Cmetal}! The gapless charge-$2$ excitations disappear for a half-fill band system when ${3J_{\vec{k}}\over4}>2|\xi_{\vec{k}}|$ $\forall \vec{k}$. In this case the $n_{\vec{k}}=1/2$ region covers the whole band and the system becomes a ($k$-Space) Valence Bond Solid (VBS) insulator.

           \subsubsection{ Hubbard Model in One-dimension}

           The Bethe-Ansatz solution of 1D Hubbard model assumes as a starting point that the number of available momentum points $N$ is equal to the number of fermions in the system\cite{hu1d}. The values of momentum $\vec{k}_i$'s are determined from a equation of the same form as Eq.\ (\ref{kequation3}), with pairwise contribution given by
         \begin{equation}
        \label{sck2}
        \kappa_{i}(\{\mathbf{K};\mathbf{Q}\}_i;U)={1\over L}\sum_{l=1}^{\bar{M}}\kappa^{cs}_{k_i;q_l}(U),
         \end{equation}
         where  $q_l$ ($l=1,..,\bar{M}$) are quantum number characterizing the spin excitations (rapidities). There is no direct exclusion statistics effect between charge momenta $k_i$'s. The spin quantum number $q_l$'s are determined by another equation
         \begin{equation}
        \label{sck3}
        0=q_l^{(0)}+{1\over L}\sum_{m=1 ((m\neq l)}^{\bar{M}}\kappa^{ss}_{q_l;q_m}(U)
         +{1\over L}\sum_{i=1}^{N}\kappa^{sc}_{q_l;k_i}(U),
         \end{equation}
         where $q_i^{(0)}=2\pi i/L$ specifies the quantum spin state. $\kappa^{ss}$ and $\kappa^{sc}$ represents exclusion statistics effects between spin excitations, and between spin and charge excitations, respectively. Eqs.\ (\ref{sck2}) and\ (\ref{sck3}) together implies that the spin excitations carry momenta and illustrate their existence through inducing momentum changes in $\vec{k}_i$'s. We note that adiabaticity (or conservation of total momentum) implies $\kappa^{ss}_{q_l;q_m}(U)=-\kappa^{ss}_{q_m;q_l}(U)$ and $\kappa^{cs}_{k_i;q_l}(U)=-\kappa^{sc}_{q_l;k_i}(U)$. The $\kappa$ functions are given by\cite{ex2, hu1d}
      \begin{eqnarray}
      \label{hub1}
       \kappa^{cs}_{k_i;q_l}(U) & = & -2\tan^{-1}({4\over U}\sin k_i-2q_l),  \\ \nonumber
       \kappa^{ss}_{q_m;q_l} & = & 2\tan^{-1}(q_i-q_j).
       \end{eqnarray}
        It was found that solution to the above equations exists for all values of $U\neq0$ at low energy with no non-analytic behavior at any finite $U$, indicating that the system is always in a state of spin-charge separation, and there is no quantum phase transition between the Fermi liquid state and the Bethe Ansatz state. The total momentum of the system is given by
       \begin{equation}
       \label{hupt}
       P=\sum_ik_i=\sum_{i=1}^Nk_i^{(0)}+\sum_{j=1}^Mq_j^{(0)},
       \end{equation}
        and is independent of interaction $U$, consistent with the adiabatic argument. The difference between the Bethe Ansatz state and the usual Fermi liquid state is that the Bethe Ansatz state is adiabatically connected to a state with total spin-charge separation, but not to the non-interacting fermion state. To see this we go to the $U\rightarrow\infty$ limit, where it is easy to see that
       \begin{equation}
       \label{hum}
         k_i\rightarrow k_i^{(0)}+{1\over L}\sum_{l=1}^{\bar{M}}2\tan^{-1}(q_l).
       \end{equation}
         We note that the correction to momentum is independent of $k_i$ and a function of the spin quantum state $\{\mathbf{Q}\}$ only. The spin rapidities $q_l$ are determined by Eq.\ (\ref{sck3}) which in the $U\rightarrow\infty$ limit is independent of the particle momentum $k_i$'s. The spin rapidities contribute to the particle momentum through Eq.\ (\ref{hupt}) and Eq.\ (\ref{hum}).

        These properties of the 1D Hubbard model are in agreement with our general description of $k$-Fermi liquid states with spin-charge separation. For more detailed discussions on the properties of the 1D Hubbard model we refer the audience to refs.\cite{bethe2} and \cite{hu1d}.

      \section{Summary}
    In this paper we introduce and develop the idea of $k$-Fermi liquids. The idea is applicable in arbitrary dimensions and goes beyond 1D Bethe Ansatz solutions or Fermi liquid states with only forward scattering\cite{forward} (at $D>1$) by asking the question ``{\em what are the most general quantum many-body states we can construct if we assume that the eigenstates of the system are described by quasi-particles carrying definite momenta $\vec{k}_i$'s}? 

     Ordinary Landau Fermi liquids and exclusion statistics are found to be unified in $k$-Fermi liquid phenomenology with the help of adiabaticity. We also show with an example the role of adiabaticity in more general situations when the $k$-Fermi liquid state is not adiabatically connected to the free fermion state.
     To go beyond Fermi liquid theory, a mechanism of spin-charge separation  based on change in the number of available $k$-states $N_k$ is proposed for spin-$1/2$ $k$-Fermi liquids where two exactly solvable models are presented to illustrate the idea: the $k$-Space Valence Bond model and the 1D Hubbard model. The two models illustrate different aspects of the $k$-Fermi liquid phenomenology associated with spin-charge separation.

    The limitations of the theory developed here should be kept in mind. The $k$-Fermi liquid phenomenlogy is an attempt to classify/construct fermionic quantum many-body states characterized by quasi-particles carrying definite set of momenta $\{\mathbf{K}\}$. In general other kinds of eigenstates exist in many-body interacting fermion systems such as bound states of fermion pairs or collective excitations\cite{bethe2}. These states are not considered in the $k$-Fermi liquid phenomenology.

      Perhaps the biggest challenge to the $k$-Fermi liquid phenomenology is whether it is supported by microscopic theories at $D>1$?  The current picture on the large $U$-limit of Hubbard model or $t-J$ model in dimensions $D>1$ based on mean-field theory or Gutzwiller Projected wavefucntions seems to suggest that these states are strongly renormalized Fermi liquid states when away from half-filling\cite{ja,nm1,nz}. Can the strong renormalization effect be understood within the $k$-Fermi liquid framework? How about spin-charge separation or the Mott transition at half-filling?


    The $k$-Space Valence Bond model we developed in this paper can be understood as an effective Landau Fermi liquid theory with a singular Landau interaction of form
         \begin{equation}
         \label{fsingular}
          f_{\vec{k}\sigma;\vec{k}'\sigma'}\sim{J_{\vec{k}}\over8}\delta^d(\vec{k}+\vec{k}')\vec{\sigma}.\vec{\sigma}'.
         \end{equation}
    Is it possible to find an effective Landau Fermi liquid model with more realistic Landau interaction that exhibits the spin-charge separation mechanism we propose in this paper? Can the Cooper pair metal state as described in this paper forms a starting point in describing realistic disordered superconductor systems\cite{Cmetal}? These are some of the unanswered questions we shall explore in the future.


  \subsubsection{acknowledgement}

           This work is supported by HKRGC through grant No. HKUST3/CRF/13G.

  \section{Appendix: Quasi-particles in $k$-Fermi liquids}
     In this appendix we give a brief introduction to the properties of quasi-particles in $k$-Fermi liquid, using spinless fermions as example. We start by recalling that in an ordinary {\em Fermi gas} a quasi-particle with momentum $\vec{p}$ is generated by adding/removing a particle with momentum  $\vec{p}$ to/from ground state.
The added particle carries energy $\varepsilon_{\vec{p}}$ and momentum $\vec{p}$. The velocity of the quasi-particle is $\vec{v}_{\vec{p}}=\nabla_{\vec{p}}\varepsilon_{\vec{p}}$. The energy and velocity of the added quasi-particle in an interacting Fermi liquid is further renormalized by the Landau interaction.

    The creation of quasi-particles is similar for $k$-Fermi liquids except that the momentum carried by all other particles are shifted when a particle is added into the system.

     Let $\{\mathbf{K}\}=(\vec{k}_1,\vec{k}_2,...,\vec{k}_N)$ be the set of occupied momentum before the quasi-particle is added to the system, and $\{\mathbf{K}^{(1)}\}=(\vec{k}_1^{(1)},\vec{k}_2^{(1)},...,\vec{k}^{(1)}_N;\vec{p}^{(1)})=(\vec{k}_1+\delta\vec{k}_1,\vec{k}_2+\delta\vec{k}_2,...,\vec{k}_N+\delta\vec{k}_N;\vec{p}+\delta\vec{p})$ be the set of occupied momentum after the quasi-particle with momentum $\vec{p}^{(1)}=\vec{p}+\delta\vec{p}$ is added to the system. We note that (i) the set $\{\mathbf{K}\}$ and $\{\mathbf{K}^{(1)}\}$ contains different number of particles and  (ii) all momentum (including the momentum $\vec{p}$ itself) are shifted when a particle is added to the system. The momentum and energy carried by this many-particle excitation represented by the difference between $\{\mathbf{K}\}$ and $\{\mathbf{K}^{(1)}\}$ are
     \begin{subequations}
     \label{qp}
     \begin{equation}
     \label{qpmomentum}
     \vec{p}_{tot}=\vec{p}^{(1)}+\sum_{i}\delta\vec{k}_i=\vec{p}_0
     \end{equation}
     and
     \begin{equation}
     \label{qpenergy}
     \bar{\varepsilon}_{\vec{p}_{tot}}=\varepsilon_{\vec{p}^{(1)}}+\sum_{i}(\varepsilon_{\vec{k}_i+\delta\vec{k}_i}-\varepsilon_{\vec{k}_i})
     \end{equation}
     respectively. We have assumed adiabaticity in equating $\vec{p}_{tot}=\vec{p}_0$.
     \end{subequations}
     The velocity of the quasi-particle excitation is $\vec{v}_{\vec{p}_{tot}}=\nabla_{\vec{p}_{tot}}\varepsilon_{\vec{p}_{tot}}$.

        Next we determine $\delta\vec{k}$'s. Using Eq.\ (\ref{kequation}). we write
      \begin{subequations}
      \label{kequationa}
      \begin{equation}
     \vec{k}_i=\vec{k}_{0i}+\vec{\kappa}_i(\{\mathbf{K}\}_i;U),
     \end{equation}
      and
      \begin{equation}
     \vec{k}_i^{(1)}=\vec{k}_i+\delta\vec{k}_i=\vec{k}_{0i}+\vec{\kappa}_i(\{\mathbf{K}^{(1)}\}_i;U).
     \end{equation}
      for $\vec{k}_i\in\{\mathbf{K}\}$ and $\vec{k}_i^{(1)}\in\{\mathbf{K}^{(1)}\}$, respectively. 
     \end{subequations}

     To proceed further we note that we expect the induced change $\delta\vec{k}$ to be of order $1/V$ ($V=$ volume) when one particle is added to a system with finite density $n=N/V$, i.e., we expect that the difference between $\vec{\kappa}_i(\{\mathbf{K}\}_i;U)$ and $\vec{\kappa}_i(\{\mathbf{K}^{(1)}\}_i;U)$ to be of order $1/V$ and we may write
     \begin{subequations}
     \label{dk}
     \begin{equation}
     \vec{\kappa}_i(\{\mathbf{K}^{(1)}\}_i;U)=\vec{\kappa}_i(\{\mathbf{K+\delta\mathbf{K}}\}_i;U)+{1\over V}\vec{\kappa}^{(1)}_i(\{\mathbf{K}\},\vec{p}^{(1)};U),
     \end{equation}
   where the first term represents corrections coming from the shift in momentum of all particles in the system except the added particle and the second term represents the direct effect of the added particle on $\vec{\kappa}_i$. The first term can be expanded in the shift in momentum $\delta\vec{k}$'s to obtain
   \begin{eqnarray}
    \vec{\kappa}_i(\{\mathbf{K+\delta\mathbf{K}}_G\}_i;U) & = & {1\over V}\sum_{j\neq i}\delta\vec{k}_j.\nabla_{\vec{k_j}}\vec{\kappa}_i(\{\mathbf{K}_G\}_i;U)  \\ \nonumber
    & & +\delta\vec{k}_i.\nabla_{\vec{k}_i}.\nabla\vec{\kappa}_i(\{\mathbf{K}_G\}_i;U).
    \end{eqnarray}
    \end{subequations}
    Combining Eqs.\ (\ref{kequationa}) and\ (\ref{dk}), we obtain a linear equation which determines $\delta\vec{k}$ to order $1/V$. Substituting $\delta\vec{k}$ into Eq.\ (\ref{qpenergy}), we can determine the quasi-particle energy. The calculation can be extended rather straightforwardly to the case of two particles to determine the corrections to Landau interaction and to the case of spin-$1/2$ fermions.



\end{document}